\documentstyle[12pt]{article}

\newcommand{\be}{\begin{equation}}
\newcommand{\ee}{\end{equation}}
\newcommand{\bea}{\begin{eqnarray}}
\newcommand{\eea}{\end{eqnarray}}

\renewcommand{\theequation}{\arabic{section}.\arabic{equation}}

\begin{document}
\begin{titlepage}
 
 
\vspace{1in}
 
\begin{center}
\Large
{\bf Qualitative Analysis of String Cosmologies}
 
\vspace{1in}

\normalsize

\large{Andrew P. Billyard$^{1a}$, Alan A. Coley$^{1,2b}$
 \& James E. Lidsey$^{3c}$}

\normalsize
\vspace{.7in}

$^1${\em Department of Physics, \\ Dalhousie University, Halifax, NS, B3H 3J5,
Canada} \\

\vspace{.1in}

$^2${\em Department of Mathematics, Statistics and Computing Science, \\ Dalhousie University,
Halifax, NS, B3H 3J5,
Canada} \\

\vspace{.1in}

$^3${\em Astronomy Centre and Centre for Theoretical Physics, \\
University of Sussex, Brighton, BN1 9QJ, U. K.}

\end{center}
 
\vspace{1in}
 
\baselineskip=24pt
\begin{abstract}
\noindent
A qualitative analysis is presented for 
spatially flat, isotropic and homogeneous 
cosmologies derived from the 
string effective action when  the 
combined effects of a dilaton, modulus, two-form potential
and central charge deficit are included. The latter 
has significant effects 
on the qualitative dynamics. The analysis is also directly applicable 
to the anisotropic Bianchi type I cosmology.
\end{abstract}

PACS NUMBERS: 98.80.Cq, 04.50.+h, 98.80.Hw
 
\vspace{.3in}
\noindent$^a$Electronic mail: jaf@mscs.dal.ca  \\
$^b$Electronic mail: aac@mscs.dal.ca \\
$^c$Electronic mail: jlidsey@astr.cpes.susx.ac.uk
 
\end{titlepage}

\section{Introduction}

\setcounter{equation}{0}

\def\theequation{\thesection.\arabic{equation}}

There has been considerable interest recently 
in the cosmological implications of string theory. String theory 
introduces significant modifications to
the standard, hot big bang model based on conventional 
Einstein gravity and 
early universe cosmology provides one of the few environments 
where the predictions of the theory can be 
quantitatively investigated. A study of string cosmologies 
is therefore well motivated. 

The evolution of the very early universe much 
below the string scale and for string coupling much smaller 
than unity, $g_s \ll 1$, is determined by ten--dimensional 
supergravity theories  \cite{eff,gsw,mr}. 
All theories of this type contain a dilaton, a graviton 
and a two--form potential 
in the Neveu--Schwarz/Neveu--Schwarz (NS--NS) bosonic sector. 
If one considers a Kaluza--Klein 
compactification from ten dimensions onto an 
isotropic six--torus of radius $e^{\beta}$, 
the effective action is given by 
\begin{equation}
\label{NSaction}
S=\int d^4 x \sqrt{-g} e^{-\Phi} \left[ R 
+\left( \nabla \Phi \right)^2 -6 \left( \nabla \beta \right)^2
-\frac{1}{12} H_{\mu\nu\lambda} H^{\mu\nu\lambda} -2\Lambda
\right]  ,
\end{equation}
where $R$ is the Ricci curvature of the spacetime with metric $g_{\mu\nu}$,  
$g\equiv {\rm det}g_{\mu\nu}$, the dilaton field, $\Phi$, 
parametrizes the string coupling, $g_s^2 \equiv  
e^{\Phi}$, and $H_{\mu\nu\lambda} \equiv 
\partial_{[\mu} B_{\nu\lambda ]}$ is the field strength 
of the  two--form potential, $B_{\mu\nu}$. The volume of the internal 
dimensions is parametrized by the modulus field, $\beta$.
The moduli fields arising from the 
compactification of the two--form on the internal 
dimensions have been neglected \cite{lower}. 
The constant,  
$\Lambda$, is determined by the central charge deficit 
of the string theory. In principle, it may take arbitrary values  
if the string is coupled to an appropriate  conformal 
field theory. Such a term may also have an origin in terms of 
the reduction of higher degree form--fields \cite{kaloper1}. 

In four dimensions, the  three--form 
field strength is dual to a one--form:  
\begin{equation}
\label{sigma}
H^{\mu\nu\lambda} \equiv e^{\Phi} \epsilon^{\mu\nu\lambda\kappa}
\nabla_{\kappa} \sigma   ,
\end{equation}
where $\epsilon^{\mu\nu\lambda\kappa}$ is the covariantly 
constant four--form. In this dual formulation, the 
field equations can be derived from the action 
\begin{equation}
\label{sigmaaction}
S=\int d^4 x \sqrt{-g} e^{-\Phi} \left[ 
R +\left( \nabla \Phi  \right)^2  -6 \left( \nabla \beta \right)^2 
-\frac{1}{2} 
e^{2 \Phi} \left( \nabla \sigma \right)^2  -2\Lambda \right]   ,
\end{equation}
where $\sigma$ is interpreted as a pseudo--scalar `axion' field
\cite{sen}. 
It can be shown 
that the action (\ref{sigmaaction}) 
is invariant under a global ${\rm SL}(2, R)$ transformation on the 
dilaton and axion fields when $\Lambda$ vanishes \cite{sen}. 
The general Friedmann--Robertson--Walker 
(FRW) cosmologies derived from Eq. 
({\ref{sigmaaction}) with $\Lambda =0$ 
have been found by employing this symmetry \cite{clw}. 
However, the symmetry is broken 
when a 
stringy cosmological constant is present \cite{kms} and the general 
FRW solution is not known in this case. 

The purpose of the present paper is to determine  
the general structure of the phase space of solutions for the class 
of spatially flat, FRW string cosmologies derived from  
the effective action (\ref{sigmaaction}) 
when a cosmological constant is present. 
This is well motivated from a theoretical 
point of view and is also 
relevant in light of recent high redshift observations 
that indicate a vacuum energy density may be 
dominating the large--scale dynamics of the universe at the present epoch 
\cite{line}. 

The paper is organized as follows. In Section 2, the 
field equations are presented as an autonomous system of 
ordinary differential equations (ODEs). 
The combined effects of the axion, modulus and dilaton fields 
are determined for a negative and positive central charge 
deficit in Sections 3 and 4, respectively. 
This extends previous qualitative analyses where 
one or more of these terms was neglected \cite{gp,kmol,emw,kmo,bf}. A
full stability analysis is performed for all cases 
by rewriting the field equations 
in terms of a set of compactified variables.  
We conclude in Section 5 with a discussion of 
the phase portraits. 

\section{Cosmological Field Equations}

\setcounter{equation}{0}

\def\theequation{\thesection.\arabic{equation}}

The spatially flat, FRW cosmological 
field equations derived from action (\ref{sigmaaction}) are given by 
\begin{eqnarray}
\label{ns1}
2\ddot{\alpha} -2\dot{\alpha}\dot{\varphi} -\dot{\sigma}^2 e^{2\varphi +6
\alpha}  &=&0\\
\label{ns2}
2\ddot{\varphi} -\dot{\varphi}^2 -3\dot{\alpha}^2 - 6 \dot{\beta}^2 
+\frac{1}{2} 
\dot{\sigma}^2 e^{2\varphi +6 \alpha}  + 2\Lambda &=&0 \\
\label{ns3}
\ddot{\beta} -\dot{\beta} \dot{\varphi} &=&0 \\
\label{ns4}
\ddot{\sigma} +\dot{\sigma} \left( \dot{\varphi} +6\dot{\alpha} \right) &=&0 ,
\end{eqnarray}
where \begin{equation}
\varphi \equiv \Phi -3\alpha
\end{equation}
defines the `shifted' dilaton field,  $a \equiv e^{\alpha}$ is the scale 
factor of the universe 
and a dot denotes differentiation with respect to 
cosmic time, $t$. The generalized Friedmann constraint equation is 
\begin{equation}
\label{nsfriedmann}
3\dot{\alpha}^2 -\dot{\varphi}^2 +6 \dot{\beta}^2 +\frac{1}{2} \dot{\sigma}^2 
e^{2\varphi + 6 \alpha} +2\Lambda =0   .
\end{equation}

A number of exact solutions to Eqs. (\ref{ns1})--(\ref{nsfriedmann}) 
are known when one or more of the degrees of freedom are trivial. 
We now discuss those that represent  
the invariant sets of the full phase space of solutions. 
The `dilaton--vacuum' solutions, where only the dilaton field 
is dynamically important, are given by 
\begin{eqnarray}
\label{dv}
a&=& a_* |t|^{1/p_{\pm}} \nonumber \\
e^{\Phi} &=& e^{\Phi_*}
|t|^{p_{\pm} -1}   ,
\end{eqnarray}
where $p_{\pm} \equiv \pm \sqrt{3}$ and $\{ a_* , \Phi_* \}$
are arbitrary constants. There 
is a curvature singularity at $t=0$. The solution (\ref{dv}) 
forms the basis of the pre--big bang scenario, 
where a growing string 
coupling can drive an epoch of inflationary (accelerated) expansion 
\cite{pbb}. The pre--big bang phase corresponds to 
the $p=p_-$ solution over the range $t<0$ and the post--big 
bang phase to the $p=p_+$ solution for $t>0$ \cite{pbb}. 
The inflationary nature of this scenario has recently 
been questioned, however \cite{question}. 

The `dilaton--moduli--vacuum' solutions, with 
$\dot{\sigma} = \Lambda =0$, are given by  
\begin{eqnarray}
\label{dmvcosmic}
a & = & a_*\left| t \right|^{\pm h_*} \nonumber \\
e^\Phi & = &e^{\Phi_*}\left|t\right|^{\pm3h_*-1} \nonumber \\
e^\beta & =& e^{\beta_*}\left|t\right|^{\pm\sqrt{(1-3h_*^2)/6}} ,
\end{eqnarray}
where $\{ h_* ,\beta_* \}$ are constants. 
This class of solution  can also be expressed in the form 
\begin{eqnarray}
a &=& a_* \left| \frac{s}{s_*} \right|^{(1+r_{\pm})/2} \nonumber \\
e^{\Phi} &=&e^{\Phi_*} \left| \frac{s}{s_*} \right|^{r_{\pm}} 
\nonumber \\
\label{dmv}
e^{\beta} &=&e^{\beta_*} \left| \frac{s}{s_*} \right|^q  ,
\end{eqnarray}
where $\{ s_*, q , r_{\pm} \} $ are constants and 
$s\equiv \int^t dt' /a(t')$ is conformal time. 
The generalized Friedmann constraint equation  (\ref{nsfriedmann}) 
leads to the constraint $r_{\pm} = \pm \left( 3-12 q^2 \right)^{1/2}$. 

The general solution where only $\Lambda =0$
is the `dilaton--moduli--axion' solution \cite{clw}: 
\begin{eqnarray}
a&=&a_* \left| \frac{s}{s_*} \right|^{1/2}
\left[ \left| \frac{s}{s_*} \right|^r +\left| 
\frac{s}{s_*} \right|^{-r} \right]^{1/2} \nonumber \\
e^{\Phi} &=&\frac{e^{\Phi_*}}{2} \left[ 
\left| \frac{s}{s_*} \right|^r + 
\left| \frac{s}{s_*} \right|^{-r} \right] \nonumber \\
\sigma &=& \sigma_* \pm e^{-\Phi_*} 
\left[ \frac{ |s/s_*|^{-r} -|s/s_*|^r}{|s/s_*|^{-r} + 
|s/s_*|^r} \right] \nonumber \\
\label{dma}
e^{\beta} &=&e^{\beta_*} \left| \frac{s}{s_*} \right|^q   ,
\end{eqnarray}
where $\sigma_*$ is an arbitrary constant and $r \equiv 
| r_{\pm}|$. This cosmology asymptotically approaches one of the 
dilaton--moduli--vacuum solutions (\ref{dmv}) in the limits 
of high and low spacetime curvature. 
The axion field induces a smooth transition between these two 
power-law solutions and 
causes a bounce to occur.  It 
is only dynamically important for a 
short time interval when $s \approx s_*$ \cite{clw}.  
 
The solutions where only the axion 
field is trivial and $\Lambda >0$ are specific 
cases of the `rolling radii' solutions \cite{mu}: 
\begin{eqnarray}
a&=&a_* \left| {\rm tanh} (A t/2) 
\right|^m \nonumber \\
e^{-\Phi} &=&e^{-\Phi_*} \left| 
\cosh (At/2) \right|^{2k -6n} 
\left| \sinh (At/2)  \right|^{2l+6n} \nonumber \\
\label{m}
e^{\beta} &=&e^{\beta_*} \left| {\rm tanh} (At/2) \right|^n   , 
\end{eqnarray}
where $A \equiv \sqrt{2\Lambda}$ and the real numbers 
$\{ k, l, m,n \}$ satisfy the constraints 
\begin{equation}
\label{m4}
3m^2 +6n^2 =1, \quad 3m+6n =k-l , \quad 
k+l =1  .
\end{equation} 
The corresponding solutions for $\Lambda < 0$ are related to 
Eqs. (\ref{m}) by redefining $A \equiv -i\tilde{A}$. 
In this case, the range of $t$ is bounded such that 
$0< t < \pi/\tilde{A}$. 

Finally, there exists the `linear dilaton vacuum' solution 
where $\Lambda >0$ \cite{myers}. This solution is static and the dilaton 
evolves linearly with time: 
\begin{equation}
\label{ldv}
\dot{a} =0 , \qquad \dot{\beta} =0 , \qquad \Phi =\pm \sqrt{2\Lambda} t  .
\end{equation}

The field equations (\ref{ns1})--(\ref{nsfriedmann}) may 
be written as the following system of autonomous ODEs: 
\begin{eqnarray}
\label{ns1a}
\dot{h} &=&\psi^2 + h \psi -3h^2 -N  -2\Lambda \\
\label{ns2a}
\dot{\psi} &=&3h^2 +N \\
\label{ns3a}
\dot{N} &=& 2N \psi \\
\label{ns4a}
\dot{\rho} &=& -6h \rho \\
\label{nsfriedmanna}
3h^2 &-&\psi^2 +N +\frac{1}{2} \rho +2\Lambda =0   ,
\end{eqnarray}
where we have defined new variables
\begin{equation}
\label{beta}
N \equiv 6\dot{\beta}^2 , \qquad \rho \equiv \dot{\sigma}^2  e^{2\varphi 
+6\alpha} , 
\qquad \psi \equiv \dot{\varphi} , \qquad h \equiv \dot{\alpha}  .
\end{equation}
The variable $\rho$ may be interpreted as the effective 
energy density of the pseudo--scalar axion field \cite{kmo}.
It follows from Eq.  (\ref{ns2a}) that $\psi$ is a
monotonically increasing function of time 
and this implies that the equilibrium
points of the system of ODEs must be located either at zero or infinite 
values of $\psi$.
In addition, due to the existence of a monotone function,
it follows that there are no periodic or recurrent orbits
in the phase space \cite{WE, Hale}. The sets
$\Lambda =0$ and $\rho =0$ are invariant sets. In particular, 
the exact solution for
$\Lambda =0$ given  by 
Eqs. (\ref{dma}) divides the phase space and the
orbits do not cross from positive to negative $\Lambda$.

We  now proceed to consider the cases where $\Lambda < 0$ and 
$\Lambda >0$ separately. 

\section{Analysis for Negative Central Charge Deficit}

\setcounter{equation}{0}

\def\theequation{\thesection.\arabic{equation}}

\subsection{Four--dimensional Model}

In this Section we consider the phase portraits of the 
NS--NS fields for negative central charge deficit.  
In the  case where the modulus  
field is frozen, $N =0$, 
Eq. (\ref{nsfriedmanna}) may be employed 
to eliminate  the axion field's energy 
density. This reduces  
the set of Eqs.  (\ref{ns1a})--(\ref{ns4a}) 
to a two--dimensional 
system. Moreover, it follows from Eq. (\ref{nsfriedmanna}) that 
\begin{equation}
\psi^2 -2\Lambda \ge 3 h^2 \ge 0
\end{equation}
and we may therefore compactify the phase space 
by defining new variables
\begin{eqnarray}
\label{eta}
\eta \equiv \frac{\psi}{\sqrt{\psi^2 -2\Lambda}}   \\
\label{xi}
\xi \equiv \frac{\sqrt{3}h}{\sqrt{\psi^2 -2\Lambda}} 
\end{eqnarray}
and a new time variable
\begin{equation}
\label{tau}
\frac{d}{d\tau} \equiv \frac{1}{\sqrt{\psi^2 -2\Lambda}} \frac{d}{dt}  .
\end{equation}
Eqs. (\ref{ns1a}) and (\ref{ns2a}) then become:
\begin{eqnarray}
\label{system1}
\frac{d\eta}{d\tau} &=& \xi^2 \left(  1- \eta^2 \right) \\ 
\label{system2}
\frac{d\xi}{d\tau}  &=& \left( \sqrt{3} + \eta \xi \right) 
\left( 1- \xi^2 \right). 
\end{eqnarray}

The variables defined in (\ref{eta}) and (\ref{xi}) are bounded, 
$\eta^2 \le 1$ and $\xi^2 \le 1$, and it follows from Eqs. 
(\ref{system1}) and (\ref{system2}) that they are both monotonically
increasing functions.  The equilibrium points are located at $\xi^2 =
\eta^2 =1$.  The invariant sets $\rho =0$ and $\Lambda =0$ correspond
to the conditions $\xi^2 =1$ and $\eta^2 =1$, respectively.  A
stability analysis indicates that the equilibrium point $A: (\eta ,
\xi )=(1,1)$ is an attractor and the point $R:(\eta , \xi ) =( -1, -1
) $ is a repeller. The points $S_{1,2}:(1, -1)\mbox{ and } (-1,1)$ are both 
saddles. The phase portrait is given in Fig.  1  and 
is discussed in Section 5, where a physical interpretation is given.

\

\

\begin{centering} [FIGURE  1 HERE] \end{centering}

\

\

\subsection{Ten--dimensional Model}

We now consider the effect of lifting the solutions to ten dimensions by 
including the modulus field, $\beta$. This will raise the dimension 
of the phase space to three. In this case, it proves convenient 
to employ the generalized Friedmann constraint  equation 
(\ref{nsfriedmanna}) to eliminate the  
modulus field rather than the axion field. 
This equation can be written as
\begin{equation} 1-\xi^2-\kappa=\frac{N}{\psi^2-2\Lambda},
\end{equation} 
where the new variable $\kappa$ is 
defined by
\begin{equation}
\label{nskappa}
\kappa \equiv \frac{\rho}{2 ( \psi^2 -2\Lambda ) } 
\end{equation}
and satisfies $0\leq\kappa\leq 1$. 
Employing Eqs. (\ref{eta})--(\ref{tau}), we can now express
the field equations  
(\ref{ns1a})--(\ref{ns4a}) in the form:
\begin{eqnarray}
\frac{d\eta}{d\tau} &=& (1-\kappa ) \left( 1- \eta^2 \right) \label{deta}\\
\frac{d\xi}{d\tau} &=& \kappa \left( \sqrt{3} +\eta \xi \right) \label{dxi}\\
\frac{d\kappa}{d\tau} &=& -2\kappa \left[ \sqrt{3} \xi + \eta ( 1- \kappa )
		\right]. \label{dkappa}
\end{eqnarray}
We note that the invariant set $N=0$ corresponds to $\kappa=1-\xi^2$,
in which case the above system of ODEs reduces to the two-dimensional
system (\ref{system1})-(\ref{system2}).

The equilibrium points of this system of ODEs all lie on one of the two
lines of non-isolated equilibrium points (or one-dimensional equilibrium sets)
\begin{equation} 
L_\pm: \eta_0^2=1,\kappa=0   ,
\end{equation}
where $\xi$ is arbitrary.  The corresponding eigenvalues are
$\lambda_1=-2\eta_0$ and $\lambda_2=-2(\sqrt{3}\xi +\eta_0)$, and
hence these equilibrium sets are normally hyperbolic (throughout, we
shall refrain from giving the corresponding eigenvectors explicitly).
The third eigenvalue is zero since this is a set of equilibrium
points.  Thus, on the line $L_+:(\eta_0=1,\kappa=0; \xi)$ the
equilibrium points are saddles for $\xi \in [-1,-1/\sqrt{3})$ and
local sinks for $\xi \in (-1/\sqrt{3},1]$. On the line
$L_-:(\eta_0=-1,\kappa=0;\xi)$ the equilibrium points are local
sources for $\xi \in [-1,1/\sqrt{3})$ and saddles for $\xi\in
(1/\sqrt{3},1]$.  The phase portrait is given in Fig.  2. The dynamics
is very simple due to the fact that the right-hand sides of
Eqs. (\ref{deta}) and (\ref{dxi}) are positive--definite and hence
$\eta$ and $\xi$ are always monotonically increasing functions. The
curved upper boundary $\kappa=1-\xi^2$ denotes the invariant set $N=0$
and therefore corresponds to Fig.  1.

\

\

\begin{centering}[FIGURE 2 HERE: `LARGE']\end{centering}

\

\

\section{Analysis for Positive Central Charge Deficit}

\setcounter{equation}{0}

\def\theequation{\thesection.\arabic{equation}}

In the case where the central charge deficit is 
positive, the variable $\psi^2 -2\Lambda$
is no longer positive--definite and therefore can not 
be employed to normalize the system. In view of this, we choose the 
normalization
\begin{equation}
\label{epsilon}
\epsilon \equiv \left( 3h^2 +\frac{1}{2} \rho +N +2\Lambda \right)^{1/2}  .
\end{equation}
The generalized Friedmann constraint equation (\ref{nsfriedmanna}) 
now takes the simple form
\begin{equation}
\label{unity}
\frac{\psi^2}{\epsilon^2} =1
\end{equation}
and may be employed to eliminate $\psi$. 
Since by definition 
$\epsilon \ge 0$, specifying one of the roots $\psi/\epsilon
=\pm 1$ corresponds to choosing the sign of $\psi$. However, it follows 
from the definition in Eq. (\ref{beta}) that changing the sign of $\psi$ 
is related to a time reversal of the dynamics. 
In what follows, we shall consider the case $\psi/\epsilon=+1$; the case $\psi/\epsilon=-1$
is qualitatively similar.

Introducing new variables
\begin{equation}
\label{nsnewa}
\mu \equiv \frac{\sqrt{3}h}{\epsilon} , 
\qquad \nu \equiv \frac{\rho}{2\epsilon^2} , \qquad 
\lambda \equiv \frac{N}{\epsilon^2} 
\end{equation}
and a new dynamical variable
\begin{equation}
\frac{d}{dT} \equiv \frac{1}{\sqrt{3} \epsilon} \frac{d}{dt}
\end{equation}
transforms Eqs. (\ref{ns1a})--(\ref{ns4a}) to the three--dimensional 
autonomous system: 
\begin{eqnarray}
\label{s3}
\frac{d \mu}{dT} &=& \nu +\frac{\mu}{\sqrt{3}}  
\left[ 1-\mu^2 -\lambda \right] \label{dmu} \\
\label{s4}
\frac{d \nu}{dT} &=& -2 \nu \left[ \mu + 
\frac{1}{\sqrt{3}} 
\left( \lambda + \mu^2 \right) \right] \label{dnu}\\
\frac{d \lambda}{d T} &=& \frac{2}{\sqrt{3}} 
\lambda \left( 1 - \mu^2 -\lambda \right) \label{dlambda}  .
\end{eqnarray}
The phase space variables are bounded by the conditions $0 \le 
\{ \mu^2 , \nu , \lambda \} \le 1$ and also satisfy $\mu^2 +\nu
+\lambda \le 1$.  The sets $\nu=0$ and $\lambda=0$ are invariant sets
corresponding to $\rho=0$ and $\dot{\beta}=0$, respectively.  In addition,
$\mu^2+\nu+\lambda=1$ is an invariant set corresponding to
$\Lambda=0$.  We note that the right-hand side of Eq. (\ref{dlambda}) is
positive-definite and this simplifies the
dynamics considerably.

The equilibrium points of the system (\ref{dmu})-(\ref{dlambda}) 
consist of the isolated equilibrium point
\begin{equation}
C:\mu=\nu=\lambda=0
\end{equation}
and the line of non-isolated equilibrium points
\begin{equation}
V:\nu=0, \lambda=1-\mu^2 \ \ \mbox{($\mu$ arbitrary)}.
\end{equation}
The eigenvalues associated with $C$ are
$\lambda_1=1/\sqrt{3}$, $\lambda_2=2/\sqrt{3}$ and
$\lambda_3=0$. Although this isolated singular point $C$ is
non-hyperbolic, a simple analysis shows that
it is a global source.
The eigenvalues associated with $V$ are:
\begin{equation}
\lambda_1=-2\left( \mu+\frac{1}{\sqrt{3}} \right)  , 
 \quad \lambda_2=-\frac{2}{\sqrt{3}} 
\end{equation}
and the third eigenvalue is zero since $V$ is an equilibrium set.
Therefore,  on $V$ the equilibrium points are saddles for $\mu\in
[-1,-1/\sqrt{3})$ and local sinks for $\mu\in(-1/\sqrt{3},1]$.  The phase
portrait is given in Fig. 3.

\

\

\begin{centering} [FIGURE 3 HERE: `LARGE'] \end{centering}

\

\

It is also 
instructive to consider the dynamics on the boundary corresponding to 
$\lambda=0$, since the case $N=0$ is of physical interest in its own right
as a four--dimensional model.  In this case 
the ODEs reduce to the two-dimensional system:
\begin{eqnarray}
\label{2d1}
\frac{d\mu}{dT} &=& \nu+\frac{\mu}{\sqrt{3}}\left(1-\mu^2\right) \\
\label{2d2}
\frac{d\nu}{dT} &=& -2\mu\nu \left[ 1 +\frac{1}{\sqrt{3}}\mu \right].
\end{eqnarray}
The equilibrium
points and their corresponding eigenvalues are:
\begin{eqnarray}
C:& \mu=    \nu=0; & \lambda_1=\frac{1}{\sqrt{3}}, \quad \lambda_2=0 \\
S:& \mu=-1, \nu=0; & \lambda_1=-\frac{2}{\sqrt{3}},  \quad 
\lambda_2=2\left(1-\frac{1}{\sqrt{3}}                       \right) \\
A:& \mu= 1, \nu=0; & \lambda_1=-\frac{2}{\sqrt{3}}, \quad 
\lambda_2=-2\left(1+\frac{1}{\sqrt{3}}                       \right).
\end{eqnarray}
Point $C$ is a non-hyperbolic equilibrium point; however, by 
changing to polar 
coordinates we find that $C$ is a repeller with an invariant ray 
$\theta=\tan^{-1}(-\sqrt{3})$.  
The saddle $S$ and the attractor $A$ lie on the 
line $V$.  The phase portrait is given in Fig. 4.

\

\

\begin{centering} [FIGURE 4 HERE] \end{centering}

\

\

This concludes the analysis of the phase portraits
for the FRW string cosmologies containing non--trivial NS--NS fields.

\section{Discussion}

\setcounter{equation}{0}

\def\theequation{\thesection.\arabic{equation}}

The phase portraits have a number of interesting features. 
In Fig. 1 the modulus field is frozen and the universe contracts from
a singular initial state. The orbits in the vicinity of the
equilibrium point $R$ are asymptotic to the $p=p_-$ dilaton--vacuum
solution (\ref{dv}).  The axion is negligible and the kinetic energy
of the dilaton dominates the energy--momentum tensor.  As the collapse
proceeds, however, the axion becomes dynamically more important and
eventually induces a bounce.  In the case of vanishing $\Lambda$,
Eqs. (\ref{dma}) imply that the future attractor would correspond to
the $p=p_+$ dilaton--vacuum solution. However, the combined effect of
the axion and central charge is to cause the universe to evolve
towards the equilibrium point $A$, where $\dot{\alpha} \rightarrow
+\infty$, in a finite time.  This behaviour differs from that found
when no axion field is present, because in this latter case there is
no bounce \cite{kmol,mu}.

This behaviour can be understood by viewing the axion field as 
a membrane \cite{kaloper}. Since this field is 
constant on the surfaces of homogeneity, the field strength 
of the two--form potential must be  directly proportional 
to the volume form of the three--space.  
If the spatial topology of the universe is that of an isotropic 
three--torus, 
the axion field can be formally interpreted as a 
membrane wrapped around this torus \cite{kaloper}. 
As the universe collapses, the membrane resists being 
squashed into a singular point and this forces the universe 
to bounce into an expansionary phase. The 
cosmological constant then dominates the axion field 
as the latter's energy density decreases. 

The inclusion of a modulus field leads to
a line of sources and sinks for the orbits (see Fig. 2). The 
axion field is dynamically negligible in the neighbourhood 
of the equilibrium points. Moreover, a bouncing cosmology 
is no longer inevitable and there exist solutions that 
expand to infinity in a finite time. The solutions are asymptotic 
to the dilaton--moduli--vacuum solutions (\ref{dmv}) near the lines 
$L_{\pm}$. The boundary points $\xi^2 =1/3$ on these lines correspond to 
the limiting cases where $\dot{\alpha}^2 =\dot{\beta}^2$. 
These represent the isotropic, ten--dimensional cosmology 
$(\dot{\alpha} = \dot{\beta})$ 
and its dual solution $(\dot{\alpha} =-\dot{\beta})$. 
In the latter solution, the 
ten--dimensional dilaton field, $\hat{\Phi} \equiv \Phi +6\beta$, 
is constant. 

In Figs. 3 and 4, the isolated equilibrium point $C$ 
corresponds to the `linear dilaton vacuum' solution (\ref{ldv})
\cite{kmol,myers}. When 
the modulus is frozen, all trajectories evolve away from $C$ 
towards the point $A$ and approach the superinflationary  
$p=p_-$ dilaton--vacuum solution
(\ref{dv}) defined over $t<0$. Some of the orbits evolving away from $C$ 
represent contracting cosmologies and the 
effect of the axion is to reverse 
the collapse in all these cases. 
For the rolling modulus solution (Fig. 3), the orbits tend to the 
dilaton--moduli--vacuum solutions as they approach the attractors 
(the sinks on $V$). As in the case of a negative 
central charge, the critical 
value $\mu^2 =1/3$ corresponds to the case where 
$\dot{\alpha}^2=\dot{\beta}^2$. 
The other boundary of $V$ is the point $A$ representing the case 
where the kinetic energy of the modulus field vanishes. 
The qualitative behaviour of models with
$\psi<0$ is similar. 

The results presented in this paper can be directly applied 
to the class of spatially flat and homogeneous
Bianchi type I models by reinterpreting the physical meaning 
of the modulus field, $\beta$. The line 
element for the axisymmetric type I model is given by
$ds^2 =-dt^2 +h_{ab}(t)dx^a dx^b$ $(a,b=1,2,3)$, where the 
metric on the surfaces of homogeneity is defined  by  
$h_{ab} \equiv e^{2\alpha}{\rm diag}[-2\beta , \beta , \beta ]$
\cite{shepley}.
The field equations derived from action (\ref{NSaction}) for 
this background 
are formally {\em identical} to those presented in Section 2 
\cite{lidsey}. In this 
anisotropic model, however,  
the variables $\alpha$ and $\beta$ now parametrize 
the averaged scale factor and the shear parameter of the 
universe, respectively. 
It would be interesting to investigate whether the above results 
can  be employed to study the question of 
isotropization in string cosmologies. 

Finally, we observe that all of the exact solutions corresponding to
the equilibrium points of the governing autonomous systems of ODEs are
self-similar since in each case the scale factor is a power-law
function of cosmic time (see, for example, Eq. (\ref{dmvcosmic}))
\cite{Maartens}.  This result can be proven in general.  For example,
all of the equilibrium points of the system of ODEs
(\ref{deta})-(\ref{dkappa}) necessarily have $h=\mbox{constant}$
(leading to a power-law solution) or $\Lambda=0$, whence from the
exact `dilaton-moduli-axion' solution (\ref{dma}) we can see that
asymptotically the scale factor is power-law.  
This implies that exact self-similar solutions play an important r\^ole in
determining the asymptotic behaviour of string cosmologies \cite{WE}.

In conclusion, therefore, we 
have presented a qualitative analysis of the spatially 
flat, FRW string cosmologies containing non--trivial dilaton, 
axion and modulus fields together with a central charge deficit 
(stringy cosmological constant). 
In all cases, variables were found that 
led to a compactification of the phase space and this
allowed a complete stability analysis to be performed. 
Both four-- and ten--dimensional 
models were studied by including the dynamics of the modulus field. 
The combined effects of 
the axion field and central charge deficit 
on the qualitative behaviour of the 
dilaton--moduli--vacuum solutions (\ref{dmvcosmic}) were determined. 
We found that such terms have a significant effect
on the dynamics. 

\vspace{.3in}

\centerline{\bf Acknowledgments}

\vspace{.3in}
APB is supported by Dalhousie University, AAC is supported
by the Natural Sciences and Engineering Research Council
of Canada (NSERC), and
JEL is supported by the Royal Society.

\vspace{1in}

\centerline{\bf Figures}

\

\noindent Figure 1: The phase portrait of the system
(\ref{system1})--(\ref{system2}) corresponding to the
four--dimensional NS--NS model with no modulus field and negative
central charge deficit $(\Lambda<0)$.  Equilibrium points are denoted
by dots and the labels in all figures correspond to those equilibrium
points (and hence the exact solutions they represent) discussed in the
text.  We shall adopt the convention throughout that large black dots
represent sources (i.e., repellers), large grey-filled dots represent
sinks (i.e., attractors), and small black dots represent saddles.  Arrows
on the trajectories have been suppressed since the direction of
increasing time is clear using this notation.

\

\

\noindent Figure 2: The phase portrait of the system
(\ref{deta})--(\ref{dkappa}).  This corresponds to the
ten--dimensional NS-NS model with negative central charge deficit
$(\Lambda<0)$.  Grey lines represent typical trajectories found within
the two-dimensional invariant sets, dashed black lines are those
trajectories along the intersection of the invariant sets, and solid
black lines are typical trajectories within the full three-dimensional
phase space.  Note that $L_\pm$ denote lines of non-isolated
equilibrium points.  See also caption to Fig. 1.

\

\

\noindent Figure 3: The phase portrait of the system
(\ref{s3})--(\ref{dlambda}) corresponding to the ten--dimensional
NS-NS model with positive central charge deficit $(\Lambda>0)$.  The
root $\psi/\epsilon=+1$ of Eq. (\ref{unity}) has been chosen.  Note
that $V$ denotes a line of non-isolated equilibrium points.  See
captions to Figs. 1 and 2.

\

\

\noindent Figure 4: Phase portrait of the system
(\ref{2d1})--(\ref{2d2}) corresponding to the four--dimensional NS-NS
model with positive central charge deficit $(\Lambda>0)$.  The root
$\psi/\epsilon=+1$ of Eq. (\ref{unity}) has been chosen.  See caption
to Fig. 1.

\

\

\noindent NOTE TO EDITOR: Figs. 2 and 3 
ARE TO BE AS``LARGE'' AS POSSIBLE.  Figs. 1 
and 4 can be ``smaller''.

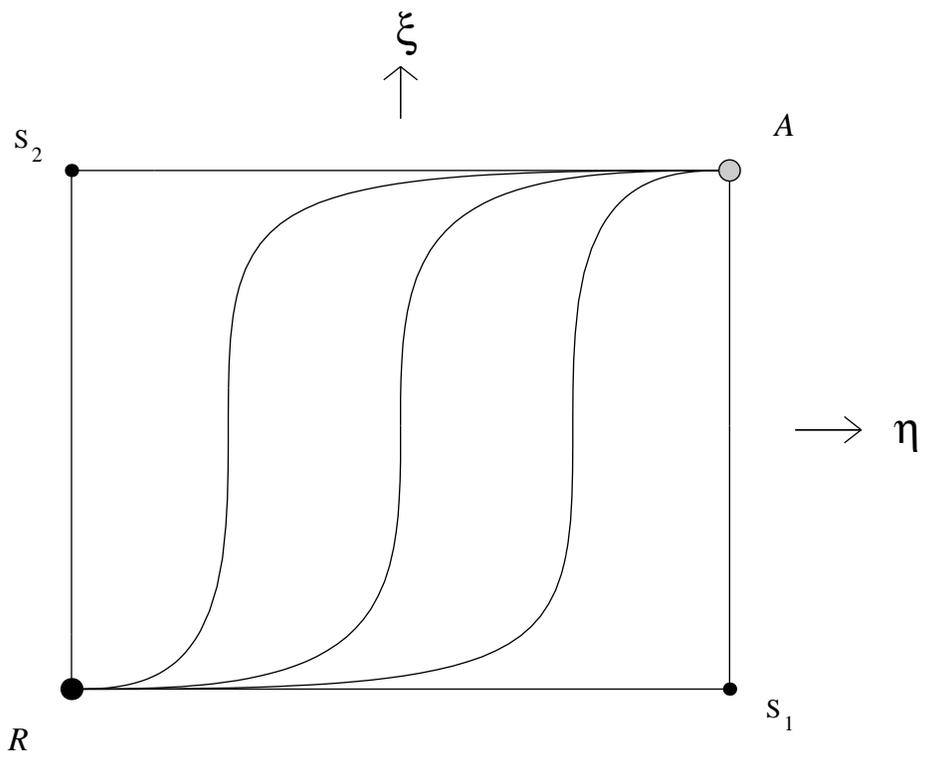
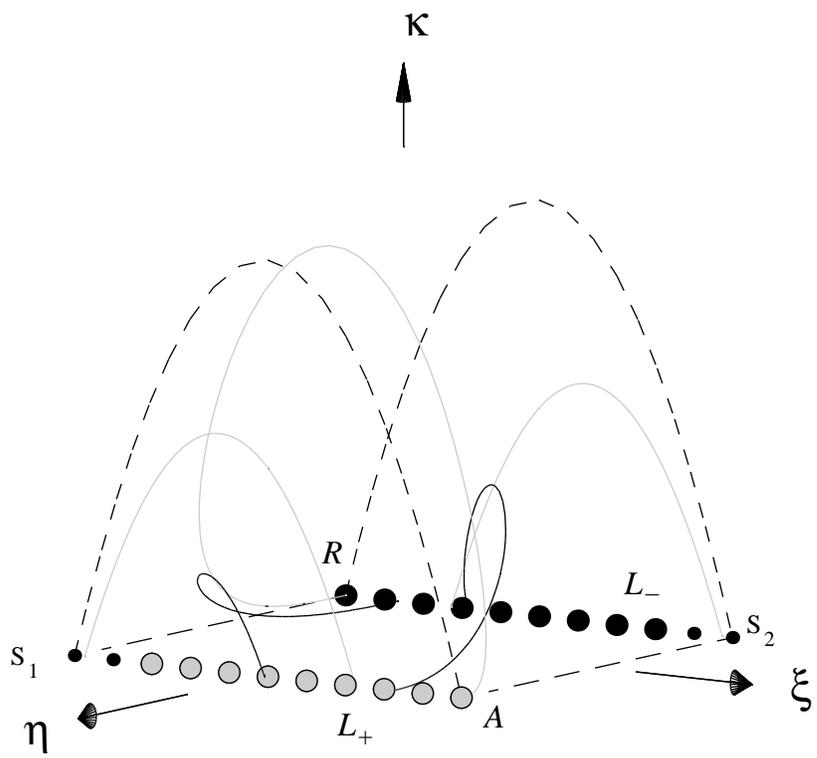
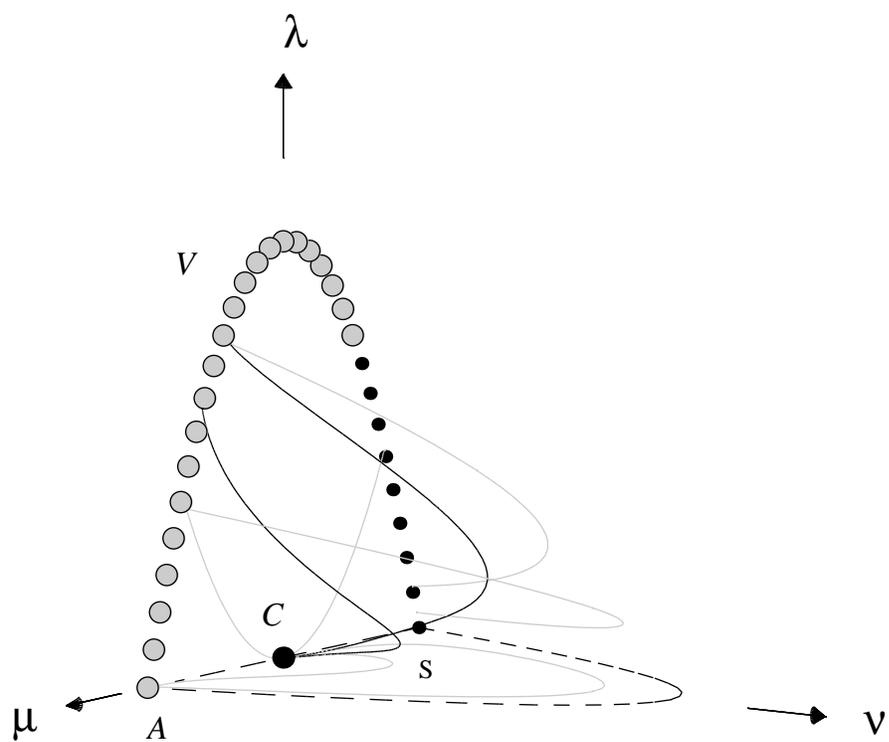
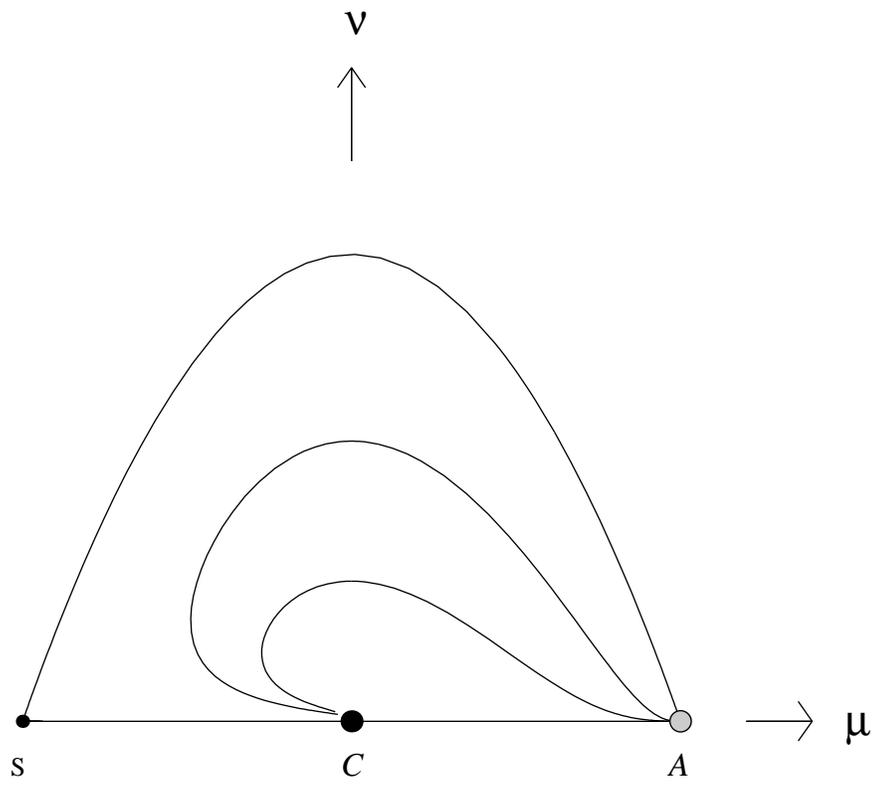
\end{document}